\documentclass[12pt,a4paper,fleqn]{article} 
\usepackage[fleqn]{amsmath} 
\usepackage{amssymb} 
\usepackage{physics} 
\usepackage{graphicx} 
\usepackage{caption} 
\usepackage[bottom]{footmisc} 
\usepackage{cite} 
\usepackage[pdftex]{hyperref} 
\newcommand{\deriv}{\mbox{d}}

\begin{document} 

\title{ 
\textbf{Time-of-flight estimation by 
utilizing Kalman filter tracking information -- \\
Part I: the concept} 
}
\author{
\textit{Winfried A. Mitaroff} \footnote{ 
\quad \href{mailto:winfried.mitaroff@oeaw.ac.at}
{\texttt{winfried.mitaroff@oeaw.ac.at} } 
} \\
Institute of High Energy Physics, \\
Austrian Academy of Sciences, Vienna 
}
\date{14 September 2021} 


\maketitle
\thispagestyle{empty}

\begin{abstract} 

Recent detector concepts at future linear or circular $e^- e^+$ colliders 
($HZ^0$ and $t \, \bar{t}$ factories) emphasize the benefits of time-of-flight 
measurements for particle identification of long-lived charged hadrons 
($\pi^{\pm}, K^{\pm}$ and $ p / \bar{p}$). 

That method relies on a precise estimation of the time-of-flight as expected, 
for a given mass hypothesis, 
from the reconstructed particle momentum and its trajectory. 
We show that for a realistic detector set-up, relativistic formulae 
are a good approximation down to lowest possible momenta. 

The optimally fitted track parameters are commonly defined near the 
interaction region. 
Extrapolation to a time-of-flight counter located behind 
the central tracking device can usually only be performed by a track model 
undisturbed from material effects. 
However, the true trajectory is distorted by multiple Coulomb scattering 
and the momentum is changed by energy loss. 
As a consequence, the estimated time-of-flight is biased 
by a large systematic error. 

This study presents a novel approach of time-of-flight estimation by 
splitting the trajectory into a chain of undisturbed track elements, 
following as close as possible the true trajectory. 
Each track element possesses an individual momentum $p_i$ 
and flight distance $l_i$. 
Remarkably, our formulae emerge by formally replacing the global 
momentum squared $p^2$ 
by the weighted harmonic mean of the individual $\{ p_i^2 \}$, 
with the weights being the corresponding individual $\{ l_i \}$. 

The optimally fitted parameters of the individual track elements can be  
obtained from track reconstruction by a Kalman filter plus smoother. 
However, care must be taken when including mass-dependent 
material effects. 

Explicit formulae for a simple scenario 
(homogeneous magnetic field and cylindrical surfaces) are given, 
together with an overview about the treatment of 
multiple Coulomb scattering and energy loss by a Kalman filter. 

A Monte Carlo study corroborating our concept will follow. 

\end{abstract} 

\newpage 

\section{Introduction} 
\label{sec:intro} 

Experiments at high-energy particle colliders consist of large complex detectors 
surrounding the beam interaction region, 
for recording the signals of particles produced in the collision events. 
A central tracking device (CTD), operating in a magnetic field, permits geometric 
reconstruction of the trajectories of long-lived charged particles 
\cite{Krammer, Fruhwirth}. 
The task of particle identification (PID) consists in determining the particle type, 
i.e. its mass \cite{Forty}. 
This may be achieved by several methods: 

Electrons $e^-$ and positrons $e^+$ are identified by electromagnetic 
calorimeters behind the CTD, and muons $\mu^\pm$ by dedicated counters 
outside the bulk of material of the detector. 
Identification of the hadrons $\pi^\pm$, $K^\pm$ and $p / \bar{p}$ are based on 
effects like specific energy loss $\deriv E / \deriv x$ in matter, 
Cherenkov or transition radiation (dependent on the velocity), 
or by measuring the time-of-flight (TOF) \cite{Forty, Klempt}. 

This study investigates only algorithmic aspects of the TOF method. 
Hardware issues like defining ``time zero'', 
achieving fast TOF counter signal responses, 
or synchronizing time stamps 
are beyond the scope of this paper. 

Section \ref{sec:tof1} presents a general overview of the TOF method, 
together with examples of calculating time differences in an ideal scenario, 
and a discussion of problems arising from the conventional approach in 
realistic scenarios. 

Section \ref{sec:tof2} introduces a novel approach by splitting the global trajectory 
into a set of smaller track elements with individual momenta, 
pinpointing to consequences arising from mass-dependence of material effects. 
This approach can be implemented 
by using information available from track fitting in the CTD 
by a Kalman filter \& smoother (KFS); 
formulae are given for a simple detector set-up. 

Section \ref{sec:tof3} gives an overview of the KFS treating with material effects. 


A summary is given in section \ref{sec:summ}. 
Appendix~\ref{sec:app1} defines the units used and some relativistic variables, 
and \ref{sec:app2} presents conventional helix tracking coordinates. 


\section{The time-of-flight method} 
\label{sec:tof1} 

The time $T$ needed for a particle of mass $m$ moving with velocity $v$ or 
momentum $p$ along a distance $L$ in the ``laboratory frame'' is given by 
\footnote{ 
See appendix \ref{sec:app1} for definitions of the variables and the units used. } 
\begin{equation} \label{equ:2-1} 
T (m) = \frac{L}{v} = \frac{L}{\beta \, c} = 
\frac{L}{c} \cdot \sqrt{1 + \eta^{-2}} = 
\frac{L}{c} \cdot \sqrt{1 + \frac{m^2}{p^2}} 
\end{equation} 
For a relativistic particle this can be approximated as 
\begin{equation} \label{equ:2-2} 
p \gg m \quad \Longrightarrow \quad 
T (m) \approx \frac{L}{c} \cdot \left( 1 + \frac{m^2}{2 \, p^2} \right) 
\end{equation} 
\indent 
First, eqs. (\ref{equ:2-1}) and (\ref{equ:2-2}) may be used to discriminate between two 
hypotheses about a particle's mass being either $m_1$ or $m_2$. 
Let $\Delta m^2 \equiv m_2^2 - m_1^2$. 
The corresponding time-of-flight difference, as a function of both $L$ and $p$, is 
\begin{equation} \label{equ:2-3} 
\Delta T = \frac{L}{c} \cdot 
\left( \sqrt{1 + \frac{m_2^2}{p^2}} - \sqrt{1 + \frac{m_1^2}{p^2}} \: \right) 
\approx \frac{L}{2 \, c \, p^2} \cdot \Delta m^2 
\end{equation} 

Second,  precise measurement of the time-of-flight $T$ by a dedicated TOF counter, 
together with accurate estimates of the flight distance $L$ and of the momentum $p$, 
yields an estimate for the particle's mass squared, viz. 
\begin{equation} \label{equ:2-4} 
m^2 = p^2 \cdot \left[ \left( \frac{c \, T}{L} \right)^2 - 1 \right] \approx 
2 \, p^2 \cdot \left( \frac{c \, T}{L} - 1 \right) 
\end{equation} 
\vspace{-4mm} 
\begin{equation} \label{equ:2-5} 
\sigma (m^2) = 2 \, p^2 \cdot \left( \frac{c \, T}{L} \right)^2 \cdot 
\left[ \frac{\sigma (T)}{T} \oplus \frac{\sigma (L)}{L} \right] 
\, \approx \, 2 \, p^2 \cdot \frac{c \, T}{L} \cdot 
\left[ \frac{\sigma (T)}{T} \oplus \frac{\sigma (L)}{L} \right] 
\end{equation} 
while the contribution of the error $\sigma (p) / p$ is suppressed by a factor 
$(c \, T / L - 1) \ll 1$ and therefore has been neglected in 
eq. (\ref{equ:2-5}).\footnote{ 
The dependence $L = L \, (p)$ implies an indirect contribution of $\sigma (p) / p$, 
which is determined by the track model and the detector set-up. 
For example, see eqs. (\ref{equ:2-1-1})--(\ref{equ:2-1-2}).} 

Meaningful measurements of $\Delta m^2$ require 
\begin{equation} \label{equ:2-6} 
\frac{\sigma (m^2)}{\Delta m^2} \approx \frac{\sigma (T)}{\Delta T} 
\oplus \frac{T}{L} \cdot \frac{\sigma (L)}{\Delta T} < 1 
\quad \Longrightarrow \quad 
\Delta T > \sigma (T) \oplus \frac{T}{L} \cdot \sigma (L) > \sigma (T) 
\end{equation} 
which, for a flight distance $L$, yields an upper limit 
\begin{equation} \label{equ:2-7} 
p^2 \approx \frac{L}{2 \, c} \cdot \frac{\Delta m^2}{\Delta T} 
< \frac{L}{2 \, c} \cdot \frac{\Delta m^2}{\sigma (T) \oplus \frac{T}{L} \cdot \sigma (L)} 
< \frac{L}{2 \, c} \cdot \frac{\Delta m^2}{\sigma (T)} 
\end{equation} 

As an example, for a time resolution $\sigma(T) = 10$~ps over a flight distance 
$L = 2$~m, separation of $\pi \leftrightarrow K$ or $K \leftrightarrow p$ could  
be achieved up to momenta $p < 8.65$~GeV/$c$ or 14.57~GeV/$c$, respectively, 
if neglecting the error $\sigma (L)$ on the estimated flight distance. 
In reality, these upper limits will be lower. 

Regarding eqs. (\ref{equ:2-3})--(\ref{equ:2-5}), the flight distance $L$ 
does not only depend on the geometry of the detector set-up, but on 
the shape of the reconstructed trajecory as well; 
hence, it is also a function of the fitted track parameters, and in particular 
of the momentum $p$. 
This dependence $L(p, \, \ldots)$ is calculated for a simple cylindrical detector set-up 
and a pure helix track model in subsection \ref{sec:tof1-1} below. 

\subsection{Discussion of a simple scenario} 
\label{sec:tof1-1}

In a detector set-up as outlined in appendix \ref{sec:app2},
the \textit{undisturbed trajectory} of a particle of unit charge and momentum $p$ 
moving in a homogeneous magnetic field of flux density $B$ is a helix 
of radius $r_H$ and slope $\cot \vartheta$: 
\begin{equation} \label{equ:2-1-1} 
r_H = \frac{p \cdot \sin \vartheta}{K_u \cdot B} \, , \qquad \qquad 
\mathrm{with} \quad K_u = 0.29979 \: \frac{\mathrm{GeV/}c}{\mathrm{T \cdot m}} 
\end{equation} 
\indent 
Assuming the particle originates at the centre point and is detected 
by a cylindrical TOF counter situated at radius $R$, its flight distance is 
\begin{equation} \label{equ:2-1-2} 
L = \frac{L_T}{\sin \vartheta} \, , \hspace{23mm} \mathrm{with} \quad 
L_T = 2 \, r_H \cdot \arcsin \frac{R}{2 \, r_H} 
\end{equation} 
being the distance projected onto the transversal plane.

Momentum $p$ and polar direction angle $\vartheta$ are known from 
track reconstruction in the CTD which is situated  in front of the TOF counter. 
If in addition coordinate $z_H$ of the hit in the TOF counter is measured 
with sufficient accuracy, 
\begin{equation} \label{equ:2-1-3} 
\cot \vartheta = \frac{z_H}{L_T} \qquad \Longrightarrow \qquad 
L = \sqrt{L_T^2 + z_H^2} 
\end{equation} 
\indent 
In order for the particle not curling back before reaching the TOF counter, 
its helix radius and its momentum must exceed a threshold 
\begin{equation} \label{equ:2-1-4} 
r_H > \frac{R}{2} \hspace{14mm} \Longrightarrow \qquad 
p > p_{min} = R \cdot \frac{K_u \cdot B}{2 \, \sin \vartheta} \, , 
\end{equation} 
hence, the projected flight distance is within $R < L_T < R \cdot \pi / 2$. 

The relative error of the transverse momentum $p_T = p \cdot \sin \vartheta$, 
estimated by a track fit, may be parametrized in the ``barrel region'' 
$| \cot \vartheta | \lesssim 1$ as 
quadratic addition of two terms, arising from detector resolution and 
multiple Coulomb scattering in CTD's material, respectively. 
The relative error of the momentum $p$, 
taking into account the correlation between $p_T$ and $\vartheta$, 
may approximately be parametrized as 
well. Formulae are given in \cite {Krammer, Fruhwirth} and are 
thoroughly discussed in \cite{Valentan}. 

However, those error formulae qualify only globally 
and are of little use for the track element approach of 
section \ref{sec:tof2} below. 
A realistic error treatment requires an adequate Monte Carlo study 
which is deferred until part II. 

\subsection{Example of calculating $\Delta T (p)$} 
\label{sec:tof1-2} 

Applied to the simple scenario of subsection \ref{sec:tof1-1}, 
a numerical calculation of the time-of-flight difference $\Delta T (p)$ 
for $\pi^\pm \leftrightarrow K^\pm$ 
and $K^\pm \leftrightarrow p / \bar{p}$, respectively, 
is performed by using 
both the exact and the approximate formula in eq. (\ref{equ:2-3}), 
and assuming 
\begin{itemize}
\item 
a pure helix trajectory undisturbed from any material effects, 
\item 
ideal tracking resolution (zero errors on the track parameters), 
\item 
detector geometry and magnetic field of a realistic set-up \cite{ILD-IDR}: 
\begin{itemize} 
\item
$R$ = 1.80 m (inner radius of a scintillation TOF counter), 
\item 
homogeneous solenoid magnetic field of $B$ = 3.5 T. 
\end{itemize} 
\end{itemize} 

\begin{figure}[h!t] 
\centering 
\includegraphics*[width=0.75\textwidth]{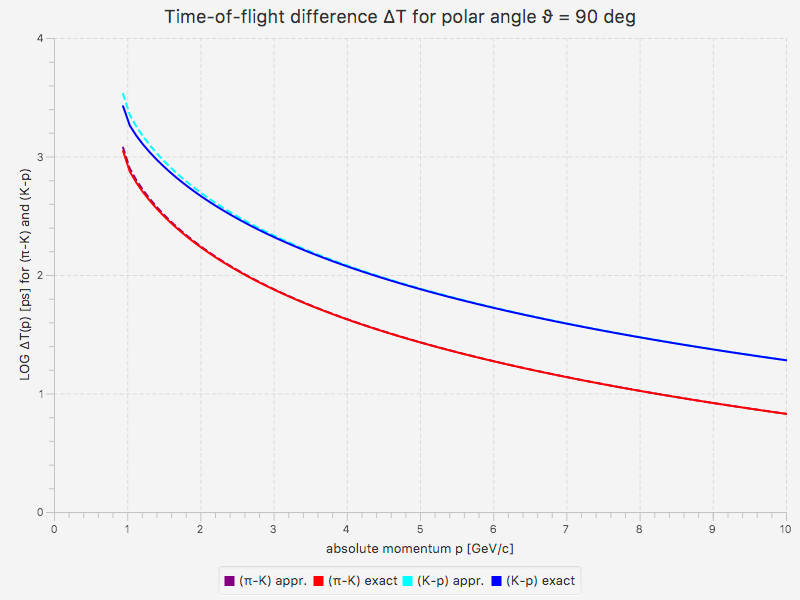} 
\caption[]{ 
$\Delta T (p)$ [ps] for polar angle $\vartheta = 90^0$; 
logarithmic ordinate.} 
\label{fig:2-2-1} 
\end{figure}

\begin{figure}[h!b] 
\centering 
\captionsetup{justification=centering} 
\includegraphics*[width=0.75\textwidth]{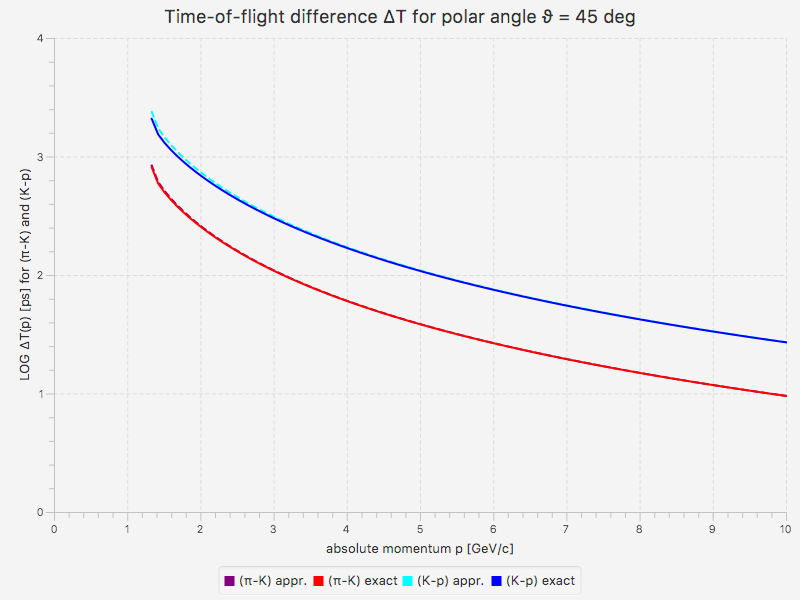} 
\caption[]{ 
$\Delta T (p)$ [ps] for polar angle $\vartheta = 45^0$;
logarithmic ordinate. \\[8mm]
Figures \ref{fig:2-2-1} and \ref{fig:2-2-2}: TOF differences for 
$\pi \leftrightarrow K$ (purple \& red) and 
$K \leftrightarrow p$ (cyan \& blue), by the exact (red/blue solid) 
and approximate (purple/cyan dashed) formula.} 
\label{fig:2-2-2} 
\end{figure}

Results are plotted in below for momenta from $p_{min}$  
(see eq. (\ref{equ:2-1-4})) up to $p = 10$~GeV/$c$, 
and for tracks at polar angles 
$\vartheta = 90^0$ (fig. \ref{fig:2-2-1}) and 
$\vartheta = 45^0$ (fig. \ref{fig:2-2-2}). 

In logarithmic scale and with the restriction 
$R < L \cdot \sin \vartheta < R \cdot \pi / 2$, 
the dependence of $\log L(p)$ on $p$ is very weak, and the curves follow 
approximately 
\begin{equation} 
\log_{10} \Delta T(p) \approx 
\mathrm{const} + \log_{10} \Delta m^2 - 2 \cdot \log_{10} p 
\end{equation} 
\indent 
The curves of the relativistic approximation can hardly be distinguished from 
those of the exact formula in eq. (\ref{equ:2-3}), 
even at the lowest possible momenta. 
This important feature will be essential for our novel approach in 
section \ref{sec:tof2} below. 

\subsection{Problems in realistic scenarios} 
\label{sec:tof1-3} 

The reconstruction of a charged particle track aims at an optimal estimate 
of the track's 5~parameters and $5 \cross 5$ covariance matrix 
at a pre-defined ``reference location''. 
This is achieved by an appropriate track model (imposed by the magnetic field) 
being fitted against the spatial measurements in the CTD, 
while taking into account any disturbance caused by material effects 
along the trajectory \cite{Krammer, Fruhwirth}. 

Since the fitted track parameters are input to a subsequent vertex fit, 
typical reference locations are e.g. the ``perigee point'' 
w.r.t. the centre of the beam interaction profile 
(see appendix \ref{sec:app2}) 
or the inner surface of the beam tube 
(in which case propagation to the perigee point is simple). 
Let's call this a ``type 1'' fit. 

Alternatively, a reference location may be chosen outside of the CTD, 
e.g. at the inner surface of the TOF counter, 
to be called ``type 2'' fit. 

Estimation of the time-of-flight requires either outward extrapolation of the 
fitted track from the perigee point to the TOF counter (type 1), 
or inward extrapolation in the opposite direction (type 2). 
In the conventional approach, both are realized by simply applying the 
track model to an undisturbed trajectory while ignoring all matter effects in between. 
Then, e.g. for a homogeneous magnetic field and a cylindrical TOF counter, 
the flight distance can be calculated by eqs. (\ref{equ:2-1-1})--(\ref{equ:2-1-3}). 

In reality, multiple Coulomb scattering causes stochastic kinks of the 2 
direction angles, and energy loss decreases the momentum $p$. 
Both effects modify the shape of the trajectory. 
As a result, the estimated TOF is expected to be biased by a systematic error 
growing larger with the detector's material budget increased. 

A recent simulation study, comparing type 1 and type 2 fits of 
$\pi^\pm$ and $K^\pm$ tracks in a realistic detector \cite{ILD-IDR}, 
found this bias changing sign but being less significant at type 2; 
the behaviour of $p / \bar{p}$ tracks was not conclusive \cite{Dudar}. 

This suggests defining a ``type~3'' fit as the \textit{weighted mean} 
\footnote{
Note that a weighted mean must be performed at the same reference location 
and must use the same set of parameters. 
This requires e.g. type 1 first to be propagated by the undisturbed track model 
to the reference location of type 2 (inner surface of the TOF counter).} 
of type 1 and type 2, then to use its undisturbed trajectory for the 
conventional approach of TOF estimation \cite{Regler}. 
We are, however, pursuing a more radical ansatz. 


\section{A novel approach to TOF} 
\label{sec:tof2}

A possible way to solve the problems discussed in subsection \ref{sec:tof1-3} above 
is based on the characteristics of modern detector set-ups. 
The material budget in the ``central region'', i.e. inside of the TOF counter, 
is concentrated in layers which are geometrically thin w.r.t. the distances between them. 
From inside out, they usually consist of the beam tube; 
silicon pixel and/or strip layers in the vertexing part of the CTD; 
followed either by many layers of an all-$Si$ tracker, 
or by a gaseous tracker (material mostly at inner and outer wall) 
optionally augmented by additional $Si$ layers \cite{Krammer, Fruhwirth}. 

This can be described by $n$ surfaces of zero geometric thickness, though 
with finite material thickness expressed in radiation lengths $X_0$ or in g/cm$^2$. 
All disturbances of the global track model by material effects 
(multiple Coulomb scattering and energy loss) 
can be accounted for by splitting the real trajectory into a set of $n + 1$ 
undisturbed track elements in front of, between, and behind those surfaces. 

Let the track elements be numbered $i = 0 \ldots n$. 
Their individual momenta $p_i$ and flight distances $l_i$ are assumed to be known 
from reconstruction in the CTD. 
As has been shown in subsection \ref{sec:tof1-2} above, 
the corresponding travelling times $t_i$ can be estimated according to 
eq. (\ref{equ:2-1}) or its relativistic approximation eq. (\ref{equ:2-2}), yielding 
\begin{equation} \label{equ:3-1} 
t_i (m) = \frac{l_i}{c} \cdot \sqrt{1 + \frac{m^2}{p_i^2}} \approx 
\frac{1}{c} \cdot \left( l_i + \frac{m^2}{2} \cdot \frac{l_i}{p_i^2} \right) 
\end{equation} 
\indent 
Total flight distance $L$ and total travelling time $T$ are 
\begin{equation} \label{equ:3-2} 
L = \sum_{i=0}^n l_i 
\end{equation} 
\vspace{-6mm} 
\begin{equation} \label{equ:3-3} 
T (m) = \sum_{i=0}^n t_i (m) \approx 
\frac{1}{c} \cdot \left( L + \frac{m^2}{2} \sum_{i=0}^n \frac{l_i}{p_i^2} \right) 
\end{equation} 
\indent
The estimated time-of-flight difference $\Delta T$ between two hypotheses, 
distinguished by the difference of the masses squared $\Delta m^2$, is given by 
\begin{equation} \label{equ:3-4} 
\Delta T \approx \frac{\Delta m^2}{2 \, c} \sum_{i=0}^n \frac{l_i}{p_i^2} 
\end{equation} 
\indent
And an estimate of the particle mass squared yields 
\begin{equation} \label{equ:3-5} 
m^2 \approx 2 \left< p^2 \right>_{HM} \cdot \left( \frac{c \, T}{L} - 1 \right) , 
\qquad \mathrm{with} \: \left< p^2 \right>_{HM} \equiv 
\sum_{i=0}^n l_i \bigg/ \sum_{i=0}^n \frac{l_i}{p_i^2} 
\end{equation} 
being the \textit{weighted harmonic mean} of $\{ \, p_i^2 \, \}$, in which 
the weights are given by the corresponding flight distances $\{ \, l_i \, \}$. 
Similar as for eq. (\ref{equ:2-5}), the contribution of the error 
$\sigma (\left< p^2 \right>_{HM})$ is suppressed by a factor 
$(c \, T / L - 1) \ll 1$, hence the error on $m^2$ is 
dominated by the measurement errors $\sigma (T)$ and $\sigma (L)$ 
alone, viz. 
\begin{equation} \label{equ:3-5bis} 
\sigma (m^2) \approx 
2 \left< p^2 \right>_{HM} \cdot \frac{c \, T}{L} \cdot 
\left[ \frac{\sigma (T)}{T} \oplus \frac{\sigma (L)}{L} \right] 
\end{equation} 
\indent 
By comparison, it is remarkable that 
eqs.~(\ref{equ:3-3})--(\ref{equ:3-5bis}) can be derived from 
eqs.~(\ref{equ:2-2})--(\ref{equ:2-5}) just by formally replacing 
\begin{equation} \label{equ:3-6} 
\frac{L}{p^2} \: \rightarrow  \: \sum_{i=0}^n \frac{l_i}{p_i^2} 
\qquad \Longrightarrow \qquad
p^2 \: \rightarrow \: \left< p^2 \right>_{HM} 
\end{equation}

\subsubsection*{Remarks about the novel approach} 

A necessary pre-requisite for this approach to work is 
knowledge of the optimally fitted track parameters 
at all $n$ surfaces as reference locations. 
This can be achieved by a Kalman filter with smoother (KFS),\footnote{ 
The resulting individual $p_i$ will be strongly correlated.} 
see subsection \ref{sec:tof2-1} below. 

If energy loss may be neglected, the momentum stays unchanged, 
i.e. $p_i = p$ for all $i$. 
In this case, only eq. (\ref{equ:3-2}) remains relevant, accounting for 
the zig-zag trajectory caused by multiple scattering; 
eqs. (\ref{equ:3-3})--(\ref{equ:3-5bis}) revert to (\ref{equ:2-2})--(\ref{equ:2-5}). 

Multiple scattering at low momenta ($\beta < 1$) and energy loss are 
dependent on the particle's mass $m$, see section \ref{sec:tof3}. 
Consequently, if those effects are to be included, 
the KFS must be performed \textit{separately for each mass hypothesis}. 
This affects above approach, eqs.~(\ref{equ:3-2})--(\ref{equ:3-5bis}): 
the $\{ \, l_i \, \}$ and $\{ \, p_i \, \}$ result from different KFS fits, 
implying that $L = \sum l_i$ and $\sum (l_i / p_i^2)$ 
become implicitly dependent on $m$, viz. 
\begin{equation} \label{equ:3-7} 
L (m)  = \left[ \sum_{i=0}^n l_i \right]_m 
\end{equation} 
\vspace{-3mm} 
\begin{equation} \label{equ:3-8} 
T (m) \approx 
\frac{1}{c} \cdot 
\left( L (m) + \frac{m^2}{2} \left[ \sum_{i=0}^n \frac{l_i}{p_i^2} \right]_m \: \right) 
\end{equation} 
\vspace{-4mm} 
\begin{equation} \label{equ:3-9} 
\Delta T = T (m_2) - T (m_1) 
\end{equation} 
replacing eqs. (\ref{equ:3-2})--(\ref{equ:3-4}). 
Alas, no simple explicit expression for an estimate of the 
particle mass squared (like in eqs. (\ref{equ:3-5})--(\ref{equ:3-5bis})) 
can be given anymore. 

\subsection{Kalman filter and smoother} 
\label{sec:tof2-1} 

The track model is in general highly non-linear, e.g. a helix to 
intersect plane or cylindrical surfaces. 
Therefore fitting is not exerted on the track parameters directly, 
but on a linear expansion of their deviation from a ``reference track'' 
to be chosen as close as possible w.r.t. the true trajectory 
in order to maintain linearity. 

State-of-the-art in track fitting is based on the Kalman filter (KF), 
a recursive and locally linear estimator.\footnote{ 
Robustification, e.g. by adaptive filters, is however non-linear.} 
Multiple scattering is regarded ``process noise'', 
but energy loss constitutes a change of the track model. 
The optimal estimate is achieved after the final filter step, 
and is equivalent to a global least squares fit. 

One asset of the KF is its ability to be supplemented by a ``smoother'' 
providing optimal track fit estimates along all previous filter steps. 
Detailed formulae for implementing a KF with smoother (KFS) are listed  
e.g. in \cite{Krammer} and \cite{Fruhwirth}. 
Explicit instructions for implementing material effects  
are given in section \ref{sec:tof3}. 

The ``reference surfaces'' defining the steps of a KFS can be classified as: 
\begin{enumerate} 
\vspace{-1mm} \item 
Active surface of zero material thickness, contributing position measurements only 
(e.g. ``virtual cylinders'' spanned by the cathode pad-rows of a TPC 
\cite{Krammer}); 
\vspace{-2mm} \item 
Passive surface of finite material thickness, without measurement information 
(beam tube, inner and outer wall of a gaseous detector, support structure, etc); 
\vspace{-2mm} \item 
Active \& passive surface combined (single- or double-sided silicon layer). 
\end{enumerate} 

Numbering the reference surfaces $S_i \, (i = 1 \ldots n)$ outward along the trajectory, 
which is represented by a chain of track elements $\wp_i$ between $S_i$ and $S_{i+1}$, 
each $\wp_i$ is defined by the smoothed parameter vector 
and its covariance matrix located at surface $S_i$ 
(in case of class 2 or 3, take the one at the ``outside face'') 
and follows the undisturbed track model. 
Adjacent elements $\wp_i^*$ ($\wp_{i-1}$ propagated to the ``inside face'' of $S_i$) 
and $\wp_i$ are connected at surface $S_i$ 
without spatial displacement, but with a kink in the direction angles 
and differing momenta in case of a class 2 or 3 surface.\footnote{ 
If energy loss can be neglected, the momenta remain equal.} 
 
Applying the KFS to the track element approach eqs. (\ref{equ:3-1})--(\ref{equ:3-5}), 
the momentum $p_i$ can be derived from the track parameters of $\wp_i$, 
and the flight distance $l_i$ can be calculated by extrapolation of  
$\wp_i \rightarrow \wp_{i+1}^*$ to surface $S_{i+1}$ 
with help of the undisturbed track model.
The final target surface $S_{n+1}$ is the TOF counter. 

The starting track element $\wp_0$ is derived from the optimal track fit 
at the perigee point. 
This is reasonable for all tracks 
originating from the primary or a short-lived secondary vertex. 
Tracks from the long-lived decays 
$K_S^0 \rightarrow \pi^+ \pi^-$ and $\Lambda^0 \rightarrow p \, \pi^-$ 
are characterized by a large transverse impact parameter 
$\left| \delta_T \right|$ and can easily be identified by their 
``V zero'' topology, hence do not require additional PID. 

\subsection{Applied to the simple scenario} 
\label{sec:tof2-2} 

Reverting to the simple scenario of subsection \ref{sec:tof1-1}, 
described by a homogeneous magnetic field $B$ 
and cylindrical reference surfaces $S_i$ at radii $R_i \, (i = 1 \ldots n + 1)$, 
modify the track model from a single undisturbed trajectory 
to one based on track elements $\wp_i \, (i = 0 \ldots n)$ 
defined by the smoothed track fit results of a KFS. 

As shown in appendix \ref{sec:app2}, the five track parameters 
(2 positions, 2 directions, and one inverse proportional to the momentum) 
can be defined by choice. Let's denote our 
smoothed parameter vector of track element $\wp_i$ at $R_i$ by 
$\wp_i = \left[ \, \Phi_i, z_i, \vartheta_i, \varphi_i, \Upsilon_i \, \right]$, 
with $\Upsilon_i = \kappa_i$ or $Q / (p_T)_i$ or $Q / p_i$. 
In case of a different parametrization provided by the KFS, 
they can be derived by an appropriate transformation. 
For the perigee parameters w.r.t. a pivot point at the origin, set 
$R_0 = \left| \delta_T \right|, \, z_0 = \delta_z$. 

Calculation of the momenta $p_i$ and flight distances $l_i$, as required by 
eqs. (\ref{equ:3-2})--(\ref{equ:3-5bis}), follows from 
eqs. (\ref{equ:2-1-1})--(\ref{equ:2-1-3}) and appendix \ref{sec:app2} as
\begin{equation} \label{equ:3-2-1} 
p_i = \frac{(p_T)_i}{\sin \vartheta_i} = 
\frac{K_u \cdot B}{\left| \kappa_i \right| \cdot \sin \vartheta_i} \, , 
\qquad \qquad 
\mathrm{with \: helix \: radius} = 1 / \left| \kappa_i \right| 
\end{equation} 
\vspace{-4mm} 
\begin{equation} \label{equ:3-2-2} 
l_i = 
\frac{\varphi_{i+1}^* - \varphi_i}{\left| \kappa_i \right| \cdot \sin \vartheta_i} 
= \sqrt{ \left( \frac{\varphi_{i+1}^* - \varphi_i}{\kappa_i} \right)^2 + 
\bigg( z_{i+1} - z_i \bigg)^2 } 
\end{equation} 
with $\varphi_{i+1}^*$ being the smoothed azimuthal direction at the 
``inside face'' of a class 2 or 3 surface $S_{i+1}$, 
otherwise $\varphi_{i+1}^* = \varphi_{i+1}$. 
Alternatively, using the cosine rule, the azimuthal directions may be 
replaced by the azimuthal positions: 
\begin{equation} \label{equ:3-2-3} 
\varphi_{i+1}^* - \varphi_i = 2 \, \arcsin \left( \frac{\left| \kappa_i \right|}{2} \cdot 
\sqrt{R_{i+1}^2 - 2 R_{i+1} R_i \cos \left( \Phi_{i+1} - \Phi_i \right) + R_i^2} 
\, \right) 
\end{equation} 
Eqs. (\ref{equ:3-2-1})--(\ref{equ:3-2-3}) will be used for the Monte Carlo study 
of part II. 

\subsubsection*{Side remark about non-helix tracking} 

In case of an inhomogeneous magnetic field requiring a track model 
based on stepwise integration by the Runge-Kutta (RK) method \cite{Fruhwirth}, 
the individual flight distances $\{ \, l_i \, \}$ are automatically calculated along 
with the integration steps. 


\section{Kalman filter with material} 
\label{sec:tof3} 

Track fitting by a Kalman filter (KF) and smoother 
\cite{Krammer, Fruhwirth} consists of the stages 
\begin{description} 
\vspace{-3mm} \item[$(A)$] 
a \textit{forward KF} running either outward or inward 
through steps $k = 1 \ldots n$; 
\vspace{-3mm} \item[$(B)$] 
a \textit{backward KF} running in the opposite direction, 
i.e. steps $k = n - 1 \ldots 1$; 
\vspace{-3mm} \item[$(C)$] 
the \textit{smoother}, performing at each reference surface 
$S_k \, (k < n)$ a weighted mean of 
$A$'s filtered estimate $\tilde{\wp}_k$ with 
$B$'s predicted estimate $\tilde{\wp}_{k,B}^{\, k+1}$, 
yielding the smoothed estimate $\tilde{\wp}_k^{\, n}$ 
which represents the optimally fitted track parameters at $S_k$. 
\end{description} 
\vspace{-2mm} 

In subsection \ref{sec:tof2-1} surfaces are labelled 
$S_i \, (i = 1 \ldots n)$ outward, hence $i = k$ in case of the 
forward KF running outward, or $i = n - k + 1$ in the other case. 
Anticipating material layers, 
the smoothed estimates $\tilde{\wp}_i^{\, n}$ are denoted 
$\wp_i^*$ or $\wp_i$ when 
located at the inside or the outside face of $S_i$, respectively 
($\wp_i^* = \wp_i$ only if surface $S_i$ is of class 1). 

Material effects are subject to the Kalman filters 
of stage $A$ and $B$. 
Formulae given below for the forward KF 
apply analogously also for the backward KF. 

The filtered estimates have large errors at early steps $k$, 
and get increasingly accurate while adding more and more 
measurements in the course of the currently running filter stage 
$A$ or $B$. 
This is a handicap concerning the accuracy of the inputs $p_k$ 
for calculating the material effects, 
i.e. multiple scattering and energy loss (see below). 

Optimal estimates at all intermediate steps $1 \ldots n$ are eventually 
available only after the smoother has been performed. 
A possible improvement can be achieved by performing the  
Kalman filter \& smoother (KFS) stages $A$--$C$ in two iterations: 

The $1^{st}$ iteration proceeds as usual. 
But the forward and backward Kalman filters of the 
$2^{nd}$~iteration now utilize 
the \textit{smoothed estimates} $\tilde{\wp}_k^{\, n}$ (momenta $p_k$) 
of the $1^{st}$~iteration as input for the calculation of 
multiple scattering and energy loss \cite{Regler}. 

The detector model of section \ref{sec:tof2} relies on layers 
of zero geometric but finite material thickness $d_k$, 
expressed in [$d$] = $X_0$ radiation lengths or in 
[$d$] = g/cm$^2$ areal mass density. 
The predicted parameters $\tilde{\wp}_k^{\, k-1}$, 
Cartesian momentum $\vec{p}_k$ and unit direction 
$\vec{u}_k = \vec{p}_k / p_k$ are defined at a position on the 
reference surface $S_k$ where the unit normal vector be $\vec{n}_k$. 
With $\alpha_k = \angle \, (\vec{u}_k, \vec{n}_k$) in space, 
the traversing length through layer $S_k$ is 
\begin{equation} \label{equ:5-3-1} 
\Delta s_k = \frac{d_k}{\cos \alpha_k} \, , \qquad \quad \mathrm{with} 
\quad \cos \alpha_k = \vec{u}_k \bullet \vec{n}_k 
\end{equation} 


In the simple scenario of subsection \ref{sec:tof2-2} with 
parameters $\tilde{\wp}_k^{\, k-1} = 
\left[ \, \Phi_k, z_k, \vartheta_k, \varphi_k, \Upsilon_k \, \right]$ 
defined at a cylindrical surface of radius $R_k$ 
(see appendix \ref{sec:app2}), 
\begin{equation} \label{equ:5-3-2} 
\cos \alpha_k = 
\left( \begin{array}{c} 
\sin \vartheta_k \: \cos \varphi_k \\ 
\sin \vartheta_k \: \sin \varphi_k \\
\cos \vartheta_k
\end{array} \right) 
\bullet 
\left( \begin{array}{c} 
\cos \Phi_k \\
\sin \Phi_k \\
0
\end{array} \right) 
= \sin \vartheta_k \cdot \cos \, (\varphi_k - \Phi_k) 
\end{equation} 
and for the parametrization 
$\tilde{\wp}_k^{\, k-1} = 
\left[ \, u_k, v_k, (t_x)_k, (t_y)_k, \Upsilon_k \, \right]$ 
defined at a plane surface $\vec{n}_k = \left[ \, 0, 0, 1 \, \right]$ 
(cf. ``case 3'' in subsections \ref{sec:tof3-1} and \ref{sec:tof3-2} below), 
\begin{equation} \label{equ:5-3-2bis} 
\cos \alpha_k = \cos \vartheta_k = 
1 \, / \sqrt{(t_x)_k^2 + (t_y)_k^2 + 1} 
\end{equation} 
both independent of the choice of the $5^{th}$ parameter 
$\Upsilon_k$. For facilitating calculations involving material effects, 
it will often be chosen as $\Upsilon_k = Q / p_k$ 
with charge $Q = \pm \, 1$. 
In case of a different parametrization by the KFS, 
care must be taken for properly transforming the parameters 
and their covariance matrix 
$\mathbb{C}_k^{\, k-1} = \mathrm{cov} ( \tilde{\wp}_k^{\, k-1} )$. 

\subsection{Multiple Coulomb scattering} 
\label{sec:tof3-1} 

Multiple Coulomb scattering off nuclei (MS) is a stochastic process 
that can be described by two projections ($\theta_{xz}, \theta_{yz}$) 
of the scattering angle $\theta$ in a local trihedron 
($z'$ pointing in the incident direction). 
Their core distributions are Gaussian,\footnote{ 
Few scatterings in extremely thin material follow a Landau distribution 
\cite{Fruhwirth}.} 
uncorrelated, with zero mean and variances 
given by the Rossi-Greisen formula \cite{Bichsel}:\footnote{ 
Correction factors like that of Highland 
are controversial and omitted here.} 
\begin{eqnarray} \label{equ:5-3-3} 
\nonumber & & \hspace{-6mm} 
\mathbb{E} (\theta_{xz}) = \mathbb{E} (\theta_{yz}) = 0 \\
& & \hspace{-6mm} 
\sigma^2 (\theta_{xz}) = \sigma^2 (\theta_{yz}) = 
\frac{\Delta s_k}{X_0} \, Q^2 \left( \frac{E_0}{\beta_k \, p_k} \right)^2, 
\qquad \mathrm{with} \quad E_0 \approx 0.0136 \: \mathrm{GeV} \\
\nonumber & & \hspace{-6mm} 
\mathrm{cov} \, (\theta_{xz}, \theta_{yz}) = 0 
\end{eqnarray} 
introducing an often overlooked mass-dependence by the factor 
$1 / \beta_k^2 = 1 + m^2 / p_k^2$, which may only be neglected 
for relativistic particles, i.e. $\beta_k \approx 1$. 

Orientation of the local trihedron ($x', y', z'$) around $z'$ is free 
and can be chosen such that $x', z'$ and global $z$ are in a plane. 
Transformation to global polar coordinates yields the 
additional variances on the direction angles caused by MS, 
\begin{equation} \label{equ:5-3-4} 
\sigma^2 (\Delta \vartheta) = \sigma^2 (\theta_{xz}), \qquad 
\sigma^2 (\Delta \varphi) \approx 
\frac{\sigma^2 (\theta_{yz})}{\sin^2 \vartheta}, \qquad 
\mathrm{cov} \, (\Delta \vartheta, \Delta \varphi) = 0 
\end{equation} 

MS is unbiased ``process noise'' of covariance 
$\mathbb{Q}_k$ in the system equation of a KF 
\cite{Krammer, Fruhwirth}. 
Let $\tilde{\wp}_k^{\, k-1}$ be the predicted parameter vector and 
$\mathbb{C}_k^{\, k-1} = \mathrm{cov} ( \tilde{\wp}_k^{\, k-1} )$ be 
its covariance at the front face of a zero-thickness passive 
(class 2 or material part of class 3) surface $S_k$. 
Then passing through $S_k$ results in an update at the rear face, 
\begin{equation} \label{equ:5-3-5} 
\begin{array}{l} 
\tilde{\wp}_k = \tilde{\wp}_k^{\, k-1} 
\\[2mm]
\mathbb{C}_k = \mathbb{C}_k^{\, k-1} + \, \mathbb{Q}_k 
\end{array} 
\end{equation} 
i.e. the parameter vector remains unchanged, whereas the covariance matrix 
is  augmented by the noise term $\mathbb{Q}_k$ 
according to the parametrization chosen. 

\subsubsection*{Case 1: \quad $\tilde{\wp}_k^{\, k-1} = 
\left[ \, \Phi_k, z_k, \vartheta_k, \varphi_k, \Upsilon_k \, \right], 
\quad \Upsilon_k = Q / p_k$} 
\begin{equation} \label{equ:5-3-5bis} 
\mathbb{Q}_k =  \mathrm{diag} 
\left[ \, 0, 0, \sigma^2 (\Delta \vartheta), \sigma^2 (\Delta \varphi), 0 \, \right] 
\end{equation} 
i.e. only the diagonal elements of the $(2 \times 2)$ sub-matrix 
of direction angles $(\vartheta, \varphi)$ are ``blown up''. 
Note that covariance $(\mathbb{Q}_k)_{3,5} = 
\mathrm{cov} (\Delta \vartheta, \Upsilon_k) = 0$. 

\subsubsection*{Case 2: \quad $\tilde{\wp}_k^{\, k-1} = 
\left[ \, \Phi_k, z_k, \vartheta_k, \varphi_k, \Upsilon'_k \, \right], 
\quad \Upsilon'_k = Q / (p_T)_k \propto \kappa_k$} 
\begin{equation} \label{equ:5-3-6} 
\mathbb{Q}_k = 
\left( \begin{array}{ccccc} 
0 & 0 & 0 & 0 & 0 \\ 
0 & 0 & 0 & 0 & 0 \\ 
0 & 0 & \sigma^2 (\Delta \vartheta) & 0 & 
- \Upsilon'_k \cot \vartheta_k \, \sigma^2 (\Delta \vartheta) \\ 
0 & 0 & 0 & \sigma^2 (\Delta \varphi) & 0 \\ 
0 & 0 & - \Upsilon'_k \cot \vartheta_k \, \sigma^2 (\Delta \vartheta) & 
0 & (\Upsilon'_k \cot \vartheta_k)^2 \, \sigma^2 (\Delta \vartheta) 
\end{array} \right) 
\end{equation} 
i.e. the matrix $\mathbb{Q}_k$ gets extra elements 
which only vanish if $\vartheta_k = \pi / 2$. 
There is maximal correlation 
$\left| \rho \left\{ (\mathbb{Q}_k)_{3,5} \right\} \right| = 
\left| \rho \left\{ \Delta \vartheta, \Upsilon'_k \right\} \right| = 1$ 
for all $\vartheta_k \neq \pi / 2$. 

\subsubsection*{Case 3: \quad $\tilde{\wp}_k^{\, k-1} = 
\left[ \, u_k, v_k, (t_x)_k, (t_y)_k, \Upsilon_k = Q / p_k \, \right]$ 
\quad \normalfont{(see appendix \ref{sec:app2})}} 
\begin{equation} \label{equ:5-3-6bis} 
\begin{array}{l} 
\mathbb{Q}_k = 
\left( \begin{array}{ccccc} 
0 & 0 & 0     & 0    & 0 \\ 
0 & 0 & 0     & 0    & 0 \\ 
0 & 0 & a_k & b_k & 0 \\ 
0 & 0 & b_k & c_k & 0 \\ 
0 & 0 & 0     & 0    & 0 
\end{array} \right), 
\qquad \mbox{with} 
\\[14mm]
\left( \begin{array}{cc} 
a_k & b_k \\[2mm] 
b_k & c_k 
\end{array} \right) \, = \, 
\left( t_x^2 + t_y^2 + 1 \right)_k 
\left( \begin{array}{cc} 
t_x^2 + 1 & t_x \, t_y \\[1mm] 
t_x \, t_y  & t_y^2 + 1 
\end{array} \right)_k 
\sigma^2 (\theta_{xz}) 
\end{array} 
\end{equation} 

\vspace{2mm} 
Note that eqs. (\ref{equ:5-3-5})--(\ref{equ:5-3-6bis}) 
apply symmetrically to the forward and backward KF, 
and there is no kink introduced by this update  step. 
The kink between the directions at either face of surface $S_k$ 
emerges only after the smoother stage. 

\subsection{Treatment of energy loss} 
\label{sec:tof3-2} 

Energy loss (EL) of a minimal ionizing particle \footnote{ 
Bremsstrahlung is negligible for particles heavier than $e^{\pm}$.} 
passing through a thin layer 
of material at $S_k$ is modelled by a discrete step $\Delta E_k$ 
of the particle's energy, 
\begin{equation} \label{equ:5-3-7} 
\Delta E_k = \Delta s_k \cdot 
\mathbb{E} \left( - \, \frac{\mathrm{d} E}{\mathrm{d} \, x} \right), 
\qquad \mathrm{with} \quad \left[ \Delta s_k \right] = \mathrm{g / cm}^2 
\end{equation} 
and the mean d$E$/d$x$ determined by the 
Bethe-Bloch formula \cite{Bichsel}: 
\begin{eqnarray} \label{equ:5-3-8} 
& & \hspace{-6mm} 
\mathbb{E} \left( - \, \frac{\mathrm{d} E}{\mathrm{d} \, x} \right) \approx  
K_0 \cdot \frac{Z}{A} \cdot \frac{Q^2}{\beta_k^2} \cdot 
\left[ \, \log \left( \frac{2 \, m_e \, \eta_k^2}{I_e} \right) - 
\beta_k^2 - \frac{1}{2} \, \delta (\eta_k) \, \right], \\[3mm] 
\nonumber & & \hspace{-6mm}  
\mbox{with} \quad K_0 \approx 
0.000307 \, \frac{\mathrm{GeV}}{\mathrm{g / cm}^2} \qquad 
\mbox{and} \quad m_e = \mbox{electron mass} 
\end{eqnarray} 
Material-dependent are $Z$ and $A$ (nuclear charge and mass number), 
$I_e$ (atom ionization energy),\footnote{ 
Take care for using the same units for both 
$\left[ I_e \right]$ and $\left[ m_e \right]$.} 
and the ``Fermi plateau'' correction 
$\delta (\eta_k)$ which may be neglected for $\eta_k \lesssim 70$. 
The Bethe-Bloch function is approximately valid for  
$0.1 \lesssim \eta_k \lesssim 1000$ 
with an accuracy $\sigma (\Delta E_k) / \Delta E_k$ of a few percent; 
its broad minimum is at $\eta_k \approx 3 \ldots 4$. 

The energy loss $\Delta E_k$ implies a discrete step of the momentum: 
\begin{eqnarray} 
\label{equ:5-3-9} & & \hspace{-6mm} 
\Delta p_k = \frac{\Delta E_k}{\beta_k} = 
\sqrt{ 1 + \frac{m^2}{p_k^2}} \cdot \Delta E_k 
\approx \left( 1 + \frac{m^2}{2 \, p_k^2} \right) \cdot \Delta E_k \\[3mm]
\label{equ:5-3-10} & & \hspace{-6mm} 
\sigma^2 (\Delta p_k) = 
\frac{\sigma^2 (\Delta E_k)}{\beta_k^2} = 
(\Delta p_k)^2 \cdot \frac{\sigma^2 (\Delta E_k)}{(\Delta E_k)^2} 
\end{eqnarray} 
which constitutes an abrupt change of the track model at surface $S_k$. 

The KF prediction at the front face of $S_k$ be 
$\tilde{\wp}_k^{\, k-1} = 
\left[ \, \Phi_k, z_k, \vartheta_k, \varphi_k, \Upsilon_k \, \right]$ 
with $\Upsilon_k = Q / p_k$ and charge $Q = \pm \, 1$ 
(``case 1''), and its covariance matrix 
$\mathbb{C}_k^{\, k-1} = \mathrm{cov} ( \tilde{\wp}_k^{\, k-1} )$. 
Passing through the material of $S_k$ yields an update at the rear face, 
with increase $+ \Delta p_k$ or decrease $- \Delta p_k$ 
if the KF running inward or outward, respectively, 
\begin{eqnarray} 
\label{equ:5-3-11} & & \hspace{-6mm} 
\tilde{\wp}_k =  
\left[ \, \Phi_k, z_k, \vartheta_k, \varphi_k, \Upsilon_k^U \, \right], 
\qquad \mbox{with} \quad \Upsilon_k^U = 
\frac{Q}{p_k \pm \Delta p_k} \\[2mm]
\label{equ:5-3-12} & & \hspace{-6mm} 
\mathbb{C}_k \approx \mathbb{C}_k^{\, k-1} +  \mathrm{diag} 
\left[ \, 0, 0, 0, 0, \frac{\sigma^2 (\Delta p_k)}{p_k^4} \, \right] 
\end{eqnarray} 
Eqs. (\ref{equ:5-3-11})--(\ref{equ:5-3-12}) apply analogously 
to ``case 3'' with $\tilde{\wp}_k^{\, k-1} = 
\left[ \, u_k, v_k, (t_x)_k, (t_y)_k, \Upsilon_k = Q / p_k \, \right]$. 
However, ``case 2'' with the $5^{th}$ parameter  
$\Upsilon'_k = Q / (p_T)_k \propto \kappa_k$ yields 
\begin{eqnarray} 
\label{equ:5-3-13} & & \hspace{-6mm} 
\tilde{\wp}_k =  
\left[ \, \Phi_k, z_k, \vartheta_k, \varphi_k, {\Upsilon'}_k^U \, \right], 
\qquad \mbox{with} \quad {\Upsilon'}_k^U = 
\frac{Q}{(p_T)_k \pm \Delta p_k \cdot \sin \vartheta} 
\\[2mm]
\label{equ:5-3-14} & & \hspace{-6mm} 
\mathbb{C}_k \approx \mathbb{C}_k^{\, k-1} +  \mathrm{diag} 
\left[ 
\, 0, 0, 0, 0, \frac{\sigma^2 (\Delta p_k)}{(p_T)_k^4} \cdot \sin^2 \vartheta 
\, \right] 
\end{eqnarray} 

All parametrizations  affect only the $5^{th}$ parameter and its variance. 
Eqs. (\ref{equ:5-3-12}) and (\ref{equ:5-3-14}) assume $\Delta p_k$ to be 
uncorrelated to $\tilde{\wp}_k^{\, k-1}$, 
and that $\Delta p_k \ll p_k$ or $(p_T)_k$. 


\section{Summary}
\label{sec:summ}

This part I of a study presents algorithmic aspects of the time-of-flight (TOF) 
method for particle identification in detectors at high-energy colliders. 

Formulae are given for discrimi\-nating between mass hypotheses or 
for the measurement of a particle's mass, 
and are applied to a pure helix trajectory in the simple scenario of a
homogeneous magnetic field and a cylindrical TOF counter. 

Plotting the numerical calculation of the TOF difference $\Delta T (p)$ between 
$\pi \leftrightarrow K$ and $K \leftrightarrow p$ as a function of the momentum 
$p$ shows the approximate formulae for relativistic particles being adequate 
even for the lowest possible momenta. 

The real particle trajectory, however, is disturbed by material effects 
(multiple Coulomb scattering and energy loss) within the detector, and 
a conventional TOF estimate using an undisturbed trajectory is biased 
by a large systematic error. 

Our novel approach to TOF estimation splits the trajectory into a chain of 
undisturbed track elements close to the true trajectory, each with an 
individual momentum $p_i$ and flight distance $l_i$ 
between discrete ``surfaces''. 
New formulae for calculating TOF estimates are given; remarkably, the 
old global $p^2$ is just replaced by the weighted harmonic mean of the 
individual $\{ p_i^2 \}$, with the weights being the individual $\{ l_i \}$. 
If energy loss may be neglected, all individual momenta $p_i$ are equal; 
nevertheless, the TOF estimate benefits from a precise calculation of the 
total flight distance $L = \sum l_i$. 

When including multiple scattering at low momenta and/or energy loss, 
which are both mass-dependent, the track elements 
must be determined separately for each mass hypothesis, 
thereby affecting above formulae of TOF estimation. 

The necessary input for this track element approach can easily be obtained 
from track reconstruction by a Kalman filter with smoother (KFS).  
Formulae are given again for the simple scenario of a homogeneous 
magnetic field and cylindrical surfaces. 
And an overview is given for 
the treatment of material effects in the KFS. 

So far, the errors on the track parameters have been ignored for TOF estimation. 
Part~II, being a follow-up to sections \ref{sec:tof2} and \ref{sec:tof3}, 
will present a Monte Carlo study based on fast simulation and reconstruction 
in a realistic detector set-up. 
It aims at quantifying 
the improvements on TOF estimation by the novel approach. 

\vspace{10mm} 

\section*{Acknowledgements}
\label{sec:ackn}

Thanks are due to \textit{Rudolf Fr\"uhwirth} and \textit{Meinhard Regler} 
(HEPHY Vienna) for helpful comments and suggestions, 
and for a careful reading of the manuscript. 


\vspace{10mm} 

\section{Appendices: formulae} 

Appendix~\ref{sec:app1} defines the units used and recalls some variables of 
relativistic kinematics, and 
appendix~\ref{sec:app2} presents conventional helix tracking coordinates 
and track parametrizations 
(updated version of a corresponding section in \cite{Krammer}). 

\subsection{Relativistic kinematics} 
\label{sec:app1}

We use a hybrid system of units, as is common practice in experimental particle physics. 
For lengths and times, SI units are used, and the vacuum speed of light is 
\begin{equation*} 
c = 2.9979 \ldots \cdot 10^8 \:  \mathrm{m / s} = 0.29979 \ldots \: \mathrm{m / ns} 
\end{equation*}
Units of energy, momentum and mass 
\footnote{
The term ``mass'' is understood to be the ``rest mass''; the notion ``relativistic mass'' 
is unnecessary, confusing, and should consistently be avoided. See also \cite{Okun}. }
are [$E$] = GeV, [$p$] = GeV/$c$ and 
[$m$] = GeV/$c^2$, respectively. 
Thus,  formulae expressed in SI units get rid of factors of $c$ by formally replacing 
$pc \rightarrow p$ and $mc^2 \rightarrow m$. 
As an example, Einstein's equation reads 
\begin{equation} \label{equ:7-1-1} 
E = \sqrt{m^2 + p^2} = \gamma \cdot m
\end{equation} 
A motion of velocity $\vec{v}$ in some inertial frame is described by the variables 
\begin{eqnarray} \label{equ:7-2} 
\nonumber & & \hspace{-6mm} 
\beta \equiv \frac{v}{c} = \sqrt{1 - \gamma^{-2}} = \frac{1}{\sqrt{1 + \eta^{-2}}} 
= \frac{p}{E} \\
& & \hspace{-6mm} 
\gamma \equiv \frac{1}{\sqrt{1 - \beta^2}} = \sqrt{1 + \eta^2} 
= \frac{E}{m} \qquad \: \: \mathrm{(Lorentz \, factor)} \\
\nonumber & & \hspace{-6mm} 
\eta \equiv \beta \cdot \gamma = \frac{\beta}{\sqrt{1 - \beta^2}} = \sqrt{\gamma^2 - 1} 
= \frac{p}{m} \qquad \mathrm{(``boost}\text{'')}   
\end{eqnarray} 
where $v, p, \beta$ and $\eta$ denote the absolute values of the corresponding vectors.


\subsection{Helix tracking coordinates} 
\label{sec:app2} 

The tracking detector layout is assumed to be approximately rotational symmetric 
w.r.t. the $z$-axis, but not necessarily mirror symmetric w.r.t. the origin $z = 0$. 
The axes $(x, y, z)$ define a right-handed orthogonal basis of 
detector-global coordinates. 
Orientation of the $x$-axis is free and may be chosen by convention. 

Surfaces are modelled in the radial (``barrel'') region as either cylinders of radius~$R$ 
or as prism planes parallel to the $z$-axis, and in the forward/backward regions 
as planes either normal to or inclined w.r.t. the $z$-axis. 
The boundary between those depends on the set-up; in $z$-symmetric detectors 
it is often around a polar angle $\vartheta \approx \pi / 4$. 
The surfaces may be purely virtual, or real layers of material. 

Besides Cartesian coordinates, cylindrical coordinates and spherical polar coordinates 
are defined for space points and/or momenta:

\vspace{2mm} 
\begin{math}
\begin{array}{l @{\hspace{10mm}} l @{\hspace{2mm}} l}
\multicolumn{3}{l}{\hspace{-12mm}
\quad \textit{Space point} \; \; \vec{x} = [ x, y, z ]_{\mathrm{cart}} 
= [ u, v, z ]_\mathrm{local} 
= [ R, \Phi, z ]_{\mathrm{cyl}} 
} \\ [2mm]
x = R \cdot \mathrm{cos} \Phi 
&
\qquad R = \sqrt{x^2 + y^2} 
\\
y = R \cdot \mathrm{sin} \Phi 
&
\qquad \Phi = \mathrm{arc tan} (y / x), 
& 
\mathrm{azimuth \; angle} \; 0 \le \Phi < \mathrm{2 \pi} 
\end{array}
\end{math}
\\ [1mm]
\indent
$\: \: u = u(x, y), \, v = v(x, y)$ are local position coordinates 
within a plane at fixed $z$.\footnote{
Local position coordinates $u(x', y'), \, v(x', y')$ may be given 
for a general plane defined by origin $\vec{x}_0$ 
and local basis vectors $(\vec{e}_1, \, \vec{e}_2)$, viz. 
$\vec{x} = \vec{x}_0 + x' \, \vec{e}_1 + y' \, \vec{e}_2$.} 

\vspace{2mm} 
\begin{math}
\begin{array}{l @{\hspace{10mm}} l @{\hspace{2mm}} l}
\multicolumn{3}{l}{\hspace{-12mm}
\quad \textit{Momentum} \; \; \vec{p} = [ p_x, p_y, p_z ]_{\mathrm{cart}} 
= [ p_T, \varphi, p_z ]_{\mathrm{cyl}} 
= [ p, \vartheta, \varphi ]_{\mathrm{sph}} 
} \\ [2mm]
p_x = p \cdot \mathrm{sin} \vartheta \cdot \mathrm{cos} \varphi 
&
p = \sqrt{p_x^2 + p_y^2 + p_z^2}  
& 
= \sqrt{p_T^2 + p_z^2}
\\
p_y = p \cdot \mathrm{sin} \vartheta \cdot \mathrm{sin} \varphi
&
\vartheta = \mathrm{arc cos} (p_z / p), 
& 
\mathrm{polar \; angle} \; 0 \le \vartheta \le \mathrm{\pi}
\\
p_T = p \cdot \mathrm{sin} \vartheta
&
\lambda \equiv \frac{\mathrm{\pi}}{2} - \vartheta,
&
\mathrm{dip \; angle} \; \frac{\mathrm{\pi}}{2} \ge \lambda \ge - \frac{\mathrm{\pi}}{2} 
\\
p_z = p \cdot \mathrm{cos} \vartheta
&
\varphi = \mathrm{arc tan} (p_y / p_x), 
& 
\mathrm{azimuth \; angle} \; 0 \le \varphi < \mathrm{2 \pi} 
\\[2mm]
\multicolumn{3}{l}{\hspace{-12mm}
\quad \mbox{In the \textit{very forward/backward direction}} 
\left| \tan \vartheta \right| \lesssim 0.5, 
\mbox{the angles (} \vartheta, \varphi \mbox{) may be replaced} }
\\
\multicolumn{3}{l}{\hspace{-12mm}
\quad \mbox{by the momentum's 
``direction tangents'' (} t_x, t_y \mbox{):}}
\\[1mm]
t_x = p_x / p_z = \tan \vartheta \cdot \cos \varphi 
&
\multicolumn{2}{l}{ 
\qquad p_x = p \cdot t_x \, / \sqrt{t_x^2 + t_y^2 + 1}} 
\\[1mm]
t_y = p_y / p_z = \tan \vartheta \cdot \sin \varphi 
&
\multicolumn{2}{l}{ 
\qquad p_y = p \cdot t_y \, / \sqrt{t_x^2 + t_y^2 + 1}} 
\\[1mm]
t_x^2 + t_y^2 \, = p_T^2 / p_z^2 = \tan^2 \vartheta 
&
\multicolumn{2}{l}{ 
\qquad p_z = \quad \; p \; / \sqrt{t_x^2 + t_y^2 + 1}} 
\\ [2mm]
\end{array}
\renewcommand{\arraystretch}{1.0}
\end{math}

The \textit{magnetic field} is assumed to be homogeneous 
and aligned parallel or antiparallel to the $z$-axis. 
It is defined by the flux density $\vec{B} = [ 0, 0, B_z ]_{\mathrm{cart}}$. 
This implies a helix track model, with the axis parallel to $z$ 
and the slope equal to $\cot \vartheta$. 

Units used are: 
[\textit{length}] = m, 
[\textit{angle}] = rad, 
[\textit{momentum}] = GeV/$c$, 
[\textit{B field}] = T, 
and [\textit{charge}] = $e$ (elementary charge).
For a particle with momentum $p$ and charge $Q$, 
the radius of the helix $r_H$ and its conveniently signed inverse $\kappa$ are 

\begin{center}
$r_H = \frac{\textstyle 1}{\textstyle K_u} \cdot 
\frac{\textstyle p \cdot \mathrm{sin} \vartheta}{\textstyle \mid Q \cdot B_z \mid} > 0$, 
\hspace{8mm}
$\kappa = - \, \mathrm{sign} (Q \cdot B_z) \, / \, r_H$ 
\end{center}

\noindent
with the unit-dependent constant (here shown for above units) 

\begin{center}
$K_u = ( 10^{-9}  \, \frac{c}{\mathrm{m / s}} ) 
\cdot \frac{\mathrm{[}length\mathrm{]}}{\mathrm{m}} \;
\frac{\mathrm{GeV}/c}{\mathrm{T} \cdot \mathrm{[}length\mathrm{]}} \, = \, 
0.29979 \ldots \frac{\mathrm{GeV}/c}{\mathrm{T} \cdot \mathrm{m}} 
$
\end{center}

\noindent 
The sign convention corresponds to 
$\mathrm{sign} (\kappa) = \mathrm{sign} ( \deriv \varphi / \deriv s ) 
\equiv $ sense of rotation in the $(x, y)$-projection. 
Note that in the absence of matter, $p$ and $\vartheta$ are constants of motion; 
in case of multiple scattering but no energy loss, $p$ remains constant.

The \textit{helix equations of motion} for a starting point $[x_S, y_S, z_S]$ 
and a starting azimuthal direction angle $\varphi_S$, 
as functions of the running parameter $\varphi$, are: 

\begin{math}
\renewcommand{\arraystretch}{1.25}
\begin{array}{l @{\hspace{4mm}} l}
x( \varphi ) = x_S + ( \mathrm{sin} \varphi - \mathrm{sin} \varphi_S ) \; / \, \kappa 
\\
y( \varphi ) = y_S - ( \mathrm{cos} \varphi - \mathrm{cos} \varphi_S ) \, / \, \kappa 
\\
z( \varphi ) = z_S + \cot \vartheta \cdot ( \varphi - \varphi_S ) \, / \, \kappa \mathrm{,} 
& 
\mathrm{path \; length} \; s( \varphi ) = ( \varphi - \varphi_S ) \, / \, ( \kappa \cdot \sin \vartheta ) 
\\ [2mm]
\end{array}
\renewcommand{\arraystretch}{1.0}
\end{math}

A track fitter's 5 \textit{internal track parameters} are defined by 2 positions, 
2 directions, and one proportional to the curvature $\kappa$,\footnote{ 
In case of zero magnetic field $\vec{B}$, i.e. straight line tracks, 
the $5^{th}$ parameter is dummy.} 
and are conveniently chosen according to the track characteristics 
and the geometry of the detector set-up, e.g. 

\begin{math}
\renewcommand{\arraystretch}{1.25}
\begin{array}{l @{\hspace{4mm}} l}
\left[ R \Phi, z, \cot \vartheta, \varphi - \Phi, \kappa \right] 
\hspace{5mm} \mbox{at} \; R = R_S, 
& 
\setminus 
\\ 
\left[ \Phi, z, \vartheta, \varphi, Q / p_T \right] 
\hspace{15mm} \mbox{at} \; R = R_S, 
& 
\mbox{ \} in the radial (``barrel'') region;} 
\\ 
\left[ \Phi, z, \vartheta, \varphi, Q / p \right] 
\hspace{17mm} \mbox{at} \; R = R_S, 
& 
/ 
\\ 
\left[ u, v, \vartheta, \varphi, Q / p \right] 
\hspace{18mm} \mbox{at} \: \: z \, = \, z_S, 
& 
\mbox{in the forward/backward regions;} 
\\
\left[ u, v, t_x, t_y, Q / p \right] 
\hspace{17mm} \mbox{at} \: \: z \, = \, z_S, 
& 
\mbox{in the very fwd./backwd. regions,}  
\\
\end{array}
\renewcommand{\arraystretch}{1.0}
\end{math}

\hspace{1mm} with $t_x = \deriv x / \deriv z, \, t_y = \deriv y / \deriv z$ 
being useful for numerical tracking along $z$.\footnote{ 
Note that $(t_x, \, t_y)$ alone cannot distinguish forward 
$(\vartheta < \pi / 2)$ vs. backward $(\vartheta > \pi / 2)$.} 
\vspace{1mm}

Examples of different \textit{external track parameters} suitably  
chosen after inward extrapolation, 
to be used as ``virtual measurements'' for a subsequent vertex fit: 

\begin{math}
\renewcommand{\arraystretch}{1.25}
\begin{array}{l @{\hspace{4mm}} l}
\left[ x, y, z, p_x, p_y, p_z \right] 
& 
\mbox{``6D Cartesian'' (note that the corresponding} 
\\ [-1mm]
& 
\hfill \qquad \quad 6 \times 6 \mbox{ covariance matrix is of rank 5 only);} 
\\
\left[ \delta_T, \delta_z, \cot \vartheta, \Phi_0, \kappa \right] 
& 
\mbox{``Perigee representation'' (see below).} 
\\ [1mm]
\end{array}
\renewcommand{\arraystretch}{1.0}
\end{math}

The \textit{perigee point} $[ x_0, y_0, z_0 ]$ of a helix track is defined, 
in the $(x,y)$-projection, 
as the point of closest approach (PCA) to a fixed ``pivot point'' $[ x_P, y_P, z_P ]$ 
which will most often be chosen at the centre of the beam interaction profile. 
The track parameters in perigee representation and usual convention are: 

\vspace{2mm} 
\begin{math}
\renewcommand{\arraystretch}{1.25}
\begin{array}{l @{\hspace{8mm}} l}
\multicolumn{2}{l}{
\delta_T = \pm \sqrt{ (x_0 - x_P)^2 + (y_0 - y_P)^2 } 
\hspace{18mm} \mbox{projected distance between the} 
}
\\ [-1mm]
&
\mbox{perigee and pivot points (transverse impact parameter),} 
\\ [-1mm]
&
\mbox{with } + \mbox{ or } - \mbox{ sign indicating the pivot point sitting to} 
\\ [-1mm]
&
\mbox{the left or to the right of the helix, respectively;} 
\\
\delta_z = z_0 - z_P
&
\mbox{distance along } z \mbox{ between perigee and pivot points;}
\\
\cot \vartheta
&
\mbox{slope of the helix;}
\\
\Phi_0 = \mathrm{arc tan} \frac{y_0 - y_P}{x_0 - x_P} 
&
\mbox{azimuthal position of perigee point  w.r.t. pivot point;} 
\\
\kappa
&
\mbox{inverse helix radius, with the sign defined as before.} 
\\ [1mm]
\end{array}
\renewcommand{\arraystretch}{1.0}
\end{math}

\noindent
Alternatively, $\Phi_0$ may be replaced by the azimuthal direction angle 
$\varphi_0 = \Phi_0 \, + \, \mathrm{sign} (\delta_T) \cdot \frac{\mathrm{\pi}}{2}$ 
of the helix at the perigee point.

Beware of subtle differences in various alternative conventions, 
e.g.\ for the units of length and momentum (affecting the value of $K_u$), 
the sign definitions for $\delta_T$ and $\kappa$, 
or a tacit assumption about $\mathrm{sign} (B_z)$ with implicit consequences. 


\newpage 
\small

\normalsize

\end{document}